\documentclass[aps,pre,onecolumn,showpacs]{revtex4}

\usepackage{graphicx,amssymb,amsmath}

\usepackage{color}

\newcommand{\be}{\begin{equation}}
\newcommand{\ee}{\end{equation}}

\newcommand{\bea}{\begin{eqnarray}}
\newcommand{\eea}{\end{eqnarray}}

\newcommand{\br}{\mathbf{r}}

\begin{document}

\title{What is the energy required to pinch a DNA plectoneme?}

\author{C\'eline Barde, Nicolas Destainville, Manoel Manghi}
\affiliation{Laboratoire de Physique Th\'eorique, IRSAMC, Universit\'e de Toulouse, CNRS \\118 route de Narbonne, F-31062 Toulouse, France, EU}

\date{\today}

\begin{abstract}
DNA supercoiling plays an important role from a biological point of view. One of its consequences at the supra-molecular level  is the formation of DNA superhelices named plectonemes. Normally separated by a distance on the order of 10 nm, the two opposite double-strands of a DNA plectoneme must be brought closer if a protein or protein complex implicated in genetic regulation is to be bound simultaneously to both strands, as if the plectoneme was locally pinched. We propose an analytic calculation of the energetic barrier, of elastic nature, required to bring closer the two loci situated on the opposed double-strands. We  examine how this energy barrier scales with the DNA supercoiling. For physically relevant values of elastic parameters and of supercoiling density, we show that the energy barrier is in the $k_{\rm B} T$ range under physiological conditions, thus demonstrating that the limiting step to loci encounter is more likely the preceding plectoneme slithering bringing the two loci side by side. 
\end{abstract}

\maketitle

\section{Introduction}

DNA supercoiling is ubiquitous in Nature and its biological role has been investigated in depth in the last decades, both experimentally and theoretically. In particular, it has been shown to facilitate the juxtaposition of sites that are distant along the DNA chain. Juxtaposition brings together in space the two sites and is required for many genetic processes such as replication, recombination or transcription~\cite{Matthews1992,Vologodskii1996,Huang2001,Lia2003}. The expression of many genes requires juxtaposition of promoters and enhancers situated on DNA loci that are non-adjacent along the chain. Enhancer-promoter interactions have been shown to be mediated by proteins bridging them specifically thanks to DNA-binding domains attaching to specific sequences of DNA, and named transcription factors (activators or repressors)~\cite{Latchman1997,Bintu2005,Nolis2009}. The Lac repressor participates to the metabolism of lactose in {\em Escherichia coli}. This transcription factor that forms a dimer bridging the two DNA double strands has been intensively studied. However, its precise mode of action is still under study~\cite{Normanno2008,Han2009,Fulcrand2016}. In eukaryotes, enhancer and promoter can be distant of up to $10^6$ base pairs (bp)~\cite{Harmston2013,Liu2016}. Interacting enhancer-promoter pairs are generically located in the same chromosome topological domain, which increases the interaction rates. It has recently been shown with the help of a mesoscopic numerical model that supercoiling of topological domains in interphase chromosomes make enhancers and promoters spend much more time in contact~\cite{Vologodskii1996,Benedetti2014}. In prokaryotes, the same kind of mechanism has been put forward, although between loci situated at more modest distances ($> 10^3$~bp) on closed circular DNA molecules. In this case as well, supercoiling has been shown to play a prominent role, both {\em in vitro}~\cite{Liu2001} and {\em in silico}~\cite{Vologodskii1996,Huang2001,Benedetti2014,Jian1998}. The typical time for loci encounter is on the order of 1 to 10~ms, even on plasmids as small as a few kbp~\cite{Huang2001,Bussiek2002}. These studies suggest that supercoiling increases the fraction of time during which the related enhancer and promoter stay together or in close proximity, even though separated by a large distance along the DNA chain. A physical mechanism at play has been proposed in Ref.~\cite{Benedetti2014}: even if the enhancer-promoter-protein(s) complex (or synapse) temporarily dissociates, the plectoneme (or superhelix) geometry ensuing from supercoiling facilitates their future re-association to form the synapse again. The average time spent in the dissociated state significantly decreases when supercoiling grows while the typical lifetime of the synapse once associated is hardly affected. 

\begin{figure}[t!]
\begin{center}
\includegraphics[height=5cm]{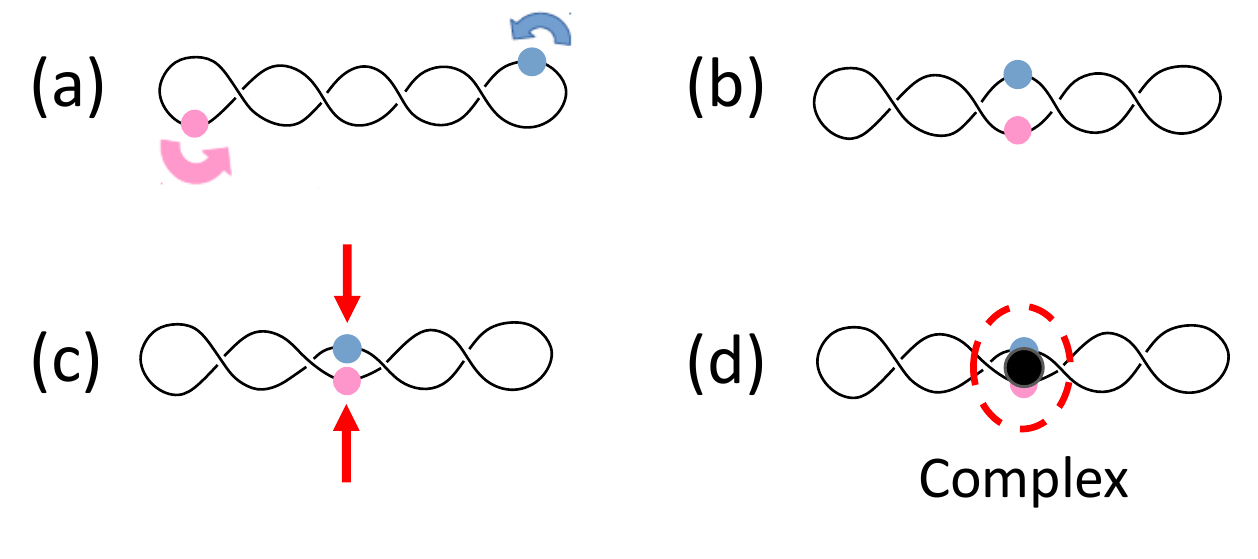}
 \caption{Sketch of plectoneme slithering and pinching. The DNA double strand is represented by a single black line (the DNA double helix is not represented). When starting from a random configuration as in (a), the two enhancer and promoter loci of interest (blue and pink dots) come at proximity in (b) by random slithering of the plectoneme, through a caterpillar-like motion. Then the plectoneme must be pinched (red arrows) to bring the enhancer and promoter sites at a sufficiently short distance called the ``capture distance'' in (c) to eventually enable the formation of the enhancer-promoter-protein(s) complex or synapse, the dashed oval in (d). Pinching has an energy cost that is under study in this work. The protein or protein complex (e.g. a transcription factor) is schematized by the black disk. 
 \label{Barde}}
 \end{center}
\end{figure}

There are two ways to ensure juxtaposition of loci distant along the chain~\cite{Huang2001}: for sufficiently long molecules ($\gtrsim 3$~kbp), the supercoiled molecule can be branched~\cite{Vologodskii1996,Fathizadeh2014} and random collisions of DNA sites that belong to different branches occur now and then. Alternatively, random slithering eventually brings the two loci on close proximity (in space) on the opposite double-strands of a DNA plectoneme (see figure~\ref{Barde}, panels (a) and (b)). We focus on the second mechanism which is the major mechanism for site juxtaposition in supercoiled DNA molecules of few kpb long under physiological conditions (notably salt conditions)~\cite{Huang2001}. The typical diameter of a DNA plectoneme is then of 10~nm. When bound in the enhancer-promoter-protein(s) complex, the two sites are a few nanometers away, and the plectoneme is thus locked by the transcription factor (see figure~\ref{Barde}, panels (c) and (d)).  

Previous studies based on numerical arguments have not explored the energy required to locally pinch the DNA plectoneme whereas the associated energy barrier might hinder the complex association and lower the association rates. In this work, we calculate analytically the energy required to pinch the plectoneme at some point. We make here an important remark: the local pinching force $\lambda$ that will be introduced below (see also figure~\ref{Barde}) is not necessarily intended to represent a real biological force ensuing from active processes. It is a convenient calculation intermediate which will enable us to compute {the elastic modulus of the plectoneme in response to local pinching. This spring constant is assumed to remain constant during the pinching process, in the frame of linear response theory. From this, we infer the work (or elastic energy)} required to deform the plectoneme and bring the two double strands closer. However this energy can in principle be brought either by any active process or by thermal fluctuations, then representing an energy barrier in Kramers' point of view~\cite{Zwanzig2001}. We will show that the pinching energy {is in} the $k_{\rm B} T$ range under physiological conditions. Therefore pinching can be achieved through thermal fluctuations alone to reach the ``capture distance'', independently of the supercoiling density. This result is in line with the conclusions of Ref.~\cite{Huang2001} that starting from a random configuration, the encounter time depends only weakly on the complex capture distance. We confirm that complex formation is a diffusion-limited stochastic process, where slithering of the plectoneme is the slow, limiting mechanism, and where hydrodynamic interactions between both strands in relative motion (figure~\ref{Barde}(a)) play a prominent role~\cite{Marko1995,Toll2001}.

Neglecting the small loops at the plectoneme extremities that however conserve the topology, the system under study in this work is made of two bendable, twistable and inextensible double-stranded-DNA (ds-DNA) molecules braided in a plectoneme, as detailed below and as illustrated in figure~\ref{fig0}. Denaturation degrees of freedom of ds-DNA are not included in the model. Indeed even though local denaturation can lead to non-linearities in case of strong bending or torsion~\cite{jpcm2009}, the deformations considered in this work remain weak. Sequence effects are also neglected at this level of modeling~\cite{Bussiek2002}. 

Following Marko and Siggia~\cite{Marko1995}, we do not fully take thermal fluctuations of the molecule shape into account. This is in part justified by the fact that the length-scales at play below are smaller than the bending and twisting persistence lengths and the molecule behaves like an elastic rigid rod at this scale. {This approximation will be tackled again in the Discussion section at the end of the article}. Entropy is only included in some effective way, as discussed below. Electrostatic interactions between different parts of the ds-DNA are not explicitly included in the model given the small value of the Debye screening length, close to 0.8~nm at physiological salt conditions. {This approximation will also be discussed at the end of this work.}

\section{The plectoneme}

The small loops at the plectoneme extremities are not taken into account in the model. However, the fact that the molecule is closed at its extremities is accounted for by the fixed linking number~\cite{Vologodskii2015}.  The plectoneme is thus modeled as a braid made of two uniform ds-DNAs of length $L$ each, $L$ being much larger than all other length scales involved in the problem, notably the persistence lengths as defined below. The double-strands are assumed to be \emph{inextensible} because the stretching modulus at physiological ionic strength  is very large, on the order of 1000~pN~$\sim100$~$k_{\rm B}T/\ell_0$ where $\ell_0 \simeq 0.34$~nm is the base-pair length~\cite{Bustamante2000}. They are modeled as ribbons defined by a curve $\mathbf{r}(s)$ setting the ds-DNA molecular axis, together with a unitary vector $\mathbf{u}(s)$ (the ribbon ``generatrix'') normal to the tangent vector $\mathbf{t}(s)$ and in the same local plane as the base-pair ``rungs'' (figure~\ref{fig0} {as well as figure~\ref{ribbon:fig} in Appendix~\ref{ribbon}}). 

\begin{figure}[ht]
\begin{center}
 \includegraphics[height=3.9cm]{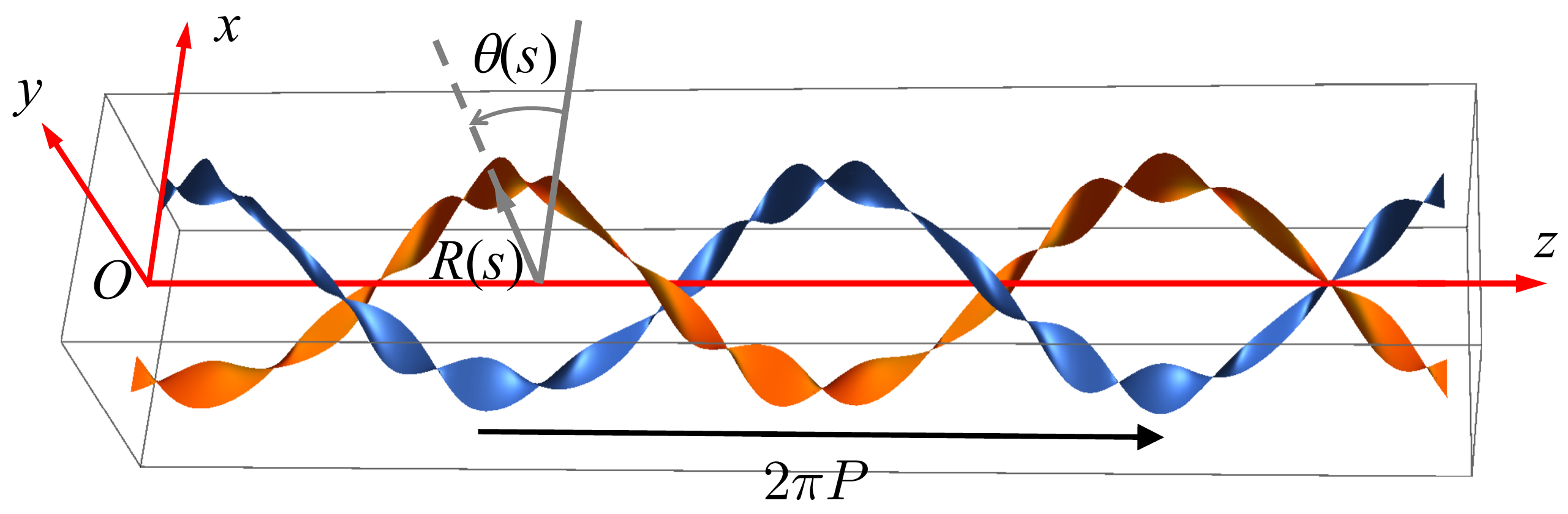}~ \includegraphics[height=3.2cm]{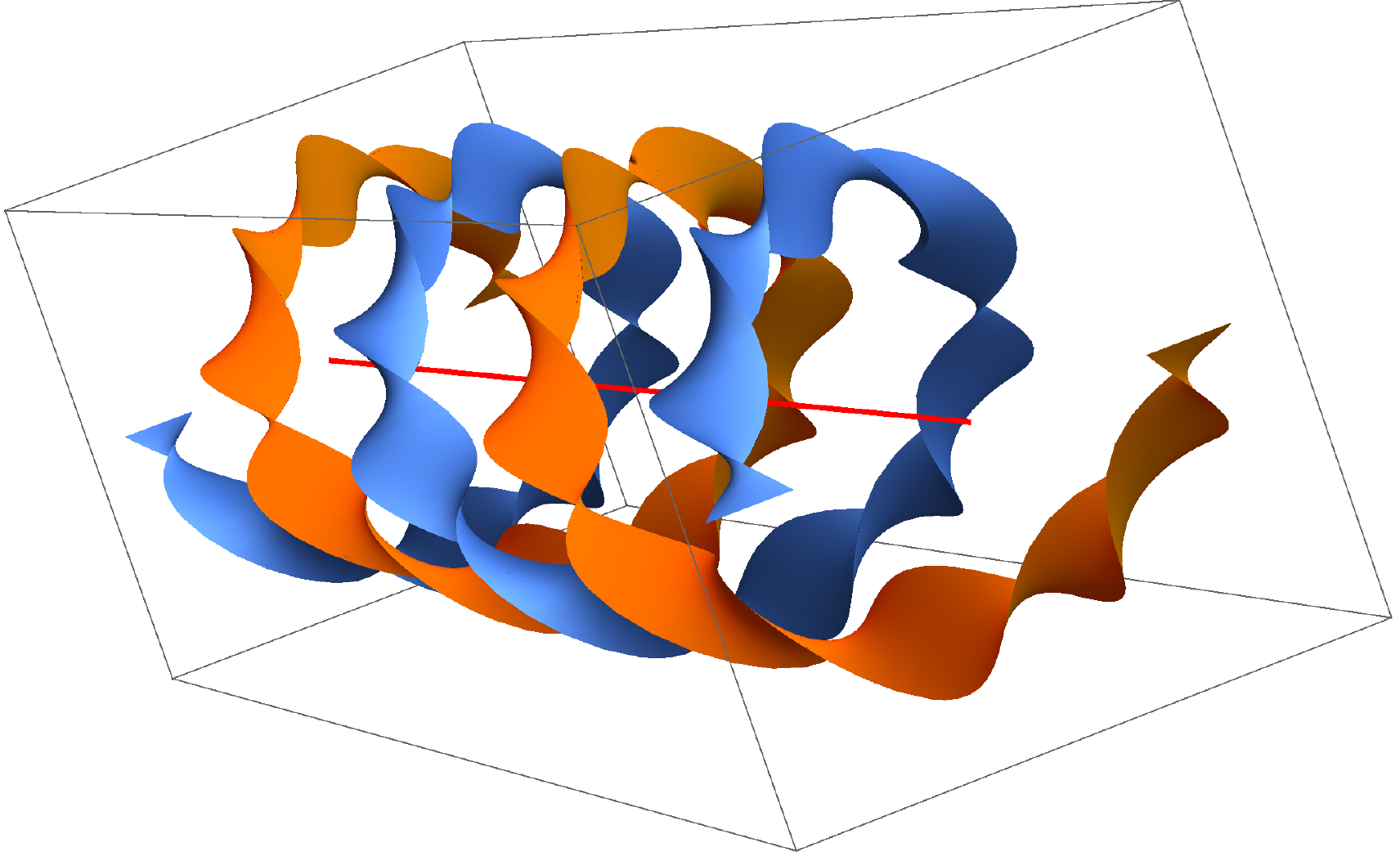}
 \caption{{Left:} Sketch of plectoneme (or superhelix) made of two ribbons, in orange and blue respectively, each of them representing two pieces of ds-DNA {forming two coaxial helices}. The plectoneme is assumed to be axi-symmetric with respect to $(Oz)$. The value of the internal pitch in the ribbons is arbitrary in this picture. The system of coordinates used in this work is illustrated in the figure. The curvilinear abscissa along the ds-DNA molecular axis is denoted by $s$. The position $\mathbf{r}(s)$ of this molecular axis is tracked by its cylindrical coordinates: the polar angle $\theta(s)$, the distance to the $z$-axis $R(s)$ and the height $z(s)$. The plectoneme pitch is $P$. {Right: same plectoneme, different point of view.} 
 \label{fig0}}
 \end{center}
\end{figure}

We suppose the problem to be axi-symmetric with respect to $(Oz)$. Hence we only consider one of the ds-DNAs, parametrized by $\mathbf{r}(s)=(x(s),y(s),z(s))$ with
\begin{eqnarray}
x(s) & = & R(s) \cos [\theta(s)] \nonumber \\
y(s) & = & R(s) \sin [\theta(s)] \label{r:of:s} \label{main:eq} \\
z(s) &  = & \int \sqrt{1 - R'^2(s) - R^2(s)[\theta'(s)]^2} \, {\rm d}s. \nonumber 
\end{eqnarray}
The polar angle (in cylindrical coordinates) is $\theta(s) \equiv \Omega(s) s$. The last coordinate $z(s)$ is imposed by the normalization of the tangent vector:
\begin{equation}
\mathbf{t} = \frac{{\rm d} \mathbf{r}}{{\rm d} s}, \quad \| \mathbf{t} \| \equiv 1.
\end{equation}
From the expressions of $x(s)$ and $y(s)$, it follows that $z'(s) = \sqrt{1 - R'^2(s) - R^2(s)[\theta'(s)]^2}$.

In the non-perturbed (non-pinched) case, $R(s) \equiv R_0$ and $\Omega(s) \equiv \Omega_0$. It ensues that $z(s)=  \sqrt{1 - R_0^2 \Omega_0^2} \, s$. Here $\Omega_0 \equiv {\rm d} \theta/ {\rm d} s$ is the angular velocity. When using the quantities usually characterizing a plectoneme, namely its radius $R_0$ and its pitch $P_0$~\cite{Marko1995} (figure~\ref{fig0}), one has 
\begin{equation}
\Omega_0 = \frac1{\sqrt{R_0^2+P_0^2}}.
\label{Om0}
\end{equation}
Note that {we have chosen in this work to call $P_0$ (and later $P$) the pitch of the plectoneme, whereas this notion of pitch sometimes refers to $2 \pi P_0$ in the literature; in addition} $\Omega_0 = \ell^{-1}$ in the notations of Ref.~\cite{Marko1995}.

Before the plectoneme is formed, the torsion density is denoted by $\sigma \omega_0$ (generally negative in bacterial DNA). After the formation of the plectoneme, it becomes $\sigma \omega_0 + \Delta \tau$~\footnote{To be completely rigorous, by convention, the {\em torsion} variation $\Delta \tau$(s), in rad/unit length, is $2\pi$ times the local {\em twist} variation $\Delta$Tw(s), in turns/unit length.} in order to minimize the elastic energy, as explained below.

\section{The free energy functional}
Assuming that the polymer elastic rod is isotropic (which is only an approximation, see, e.g., Ref.~\cite{Carlon2017}), the free energy reads in units of $k_{\rm B}T$~\cite{Marko1995,Chouaieb2006,Marko2015}:
\begin{equation}
F \equiv \frac{\ell_p}2 \int_{-L/2}^{L/2} {\rm d}s \, \left\| \frac{{\rm d} \mathbf{t}}{{\rm d} s} \right\|^2
+ \frac{C}2 \int_{-L/2}^{L/2} {\rm d}s \, [\sigma \omega_0 + \Delta \tau (s)]^2
+ K \int_{-L/2}^{L/2}  \frac{{\rm d}s}{R(s)^{2/3}}
- \lambda [R(0)-R_c].
\label{F:gen}
\end{equation}
The first two terms define the twistable worm-like chain model. Here $\ell_p \simeq 50$~nm is the bending persistence length~\cite{Brunet2015} and $C \simeq 110$~nm the torsional persistence length. The  measured values of $C$ significantly depend on the experimental technique, but it has recently been explained that this comes from the twist-bend coupling ensuing from the pronounced difference between the minor and major grooves of DNA~\cite{Carlon2017}. The intrinsic value of $C$ is close to 110~nm, but assuming an isotropic model (i.e. neglecting the twist-bend coupling) leads to a lower renormalized value of $C$ in absence of stretching forces, close to 75~nm. We shall discuss this alternative value below.

The algebraic third term takes into account the fact that both ds-DNA cannot intersect. This repulsion is accounted for in an effective way, with $K = c/\ell_p^{1/3}$ where $c$ is a dimensionless constant close to 1~\cite{Marko1995,Marko2015,Dijkstra1993,Burkhardt1995,Chen2016}. We set $c=1$ in this work. Without this entropic term, the plectoneme would collapse into a line, both double strands being superimposed with the axis $(Oz)$~\cite{Marko1995}. Finally $\lambda$ is a Lagrange multiplier (homogeneous to a pinching force) ensuring that $R(s=0)$ equals the capture radius $R_c<R_0$. As it is defined $\lambda$ must be negative to enforce $R(s=0)<R_0$. Short-ranged hard-core repulsion, notably of electrostatic origin as far as DNA is concerned~\cite{Marko1995}, is not included in the free energy at this level of modeling. {The electrostatic part has been estimated~\cite{Brahmachari2017}, taking into account both the inter-helix contribution~\cite{Ubbink1999} and the electrostatic self-energy of each helix of the plectoneme. The expression obtained in this work could in principle be inserted in equation~\eqref{F:gen}. However, it vanishes when $R_0 \gg \lambda_D$, the Debye screening length. Under physiological conditions, $\lambda_D \approx 0.8$~nm. We shall see below that for biological values of the supercoiling density, $R_0$ remains larger than $\lambda_D$, which justifies to neglect the electrostatic contribution. However, when strongly pinching the plectoneme, the local inter-strand distance can become on the order of $\lambda_D$. This point will be re-examined at the end of the paper.}

 
\subsection{Non-perturbed plectoneme ($\lambda=0$)}

In the non-perturbed, homogeneous plectoneme, one easily finds the curvature and torsion of $\mathbf{r}(s)$~\cite{Marko1995,Chouaieb2006}:
\begin{eqnarray}
\gamma_0 & = & \left\| \frac{{\rm d} \mathbf{t}}{{\rm d} s} \right\| = \frac{R_0}{R_0^2+P_0^2} = R_0 \Omega_0^2 \label{gamma0} \label{gamma0} \\
\Delta \tau_0 & = & \frac{P_0}{R_0^2+P_0^2} = P_0 \Omega_0^2 =   \Omega_0 
	\sqrt{1 - R_0^2 \Omega_0^2}. \label{Delta:tau0}
\end{eqnarray}
The torsion variation $\Delta \tau_0$ when forming the plectoneme can {be computed through the Frenet-Serret formula 
\begin{equation} 
\tau = - \frac{{\rm d} \mathbf{b}}{{\rm d}s} \cdot \mathbf{n} \quad {\rm with} \quad \mathbf{n}=\frac{{\rm d} \mathbf{t}/{\rm d} s}{\left\| {\rm d} \mathbf{t}/{\rm d} s \right\|}.
\end{equation} 
The unit vector $\mathbf{n}$ is called the normal and $\mathbf{b}=\mathbf{t} \times \mathbf{n}$ is the binormal.} Here $\Delta \tau_0 >0$ because $\sigma$ is chosen $<0$ as in generic bacterial DNA.

Note that the torsion variation $\Delta \tau_0$ when forming the plectoneme {could alternatively} be computed through the Calugare\-anu-Fuller-White theorem~\cite{Marko2015,Vologodskii2015} $\Delta{\rm Lk} = L \, \Delta \tau_0/(2\pi) + \Delta {\rm Wr}=0$, using the fact that the linking number ${\rm Lk}$ is a topological invariant {when the plectoneme is formed. Thus $\Delta \tau_0 = - 2 \pi \, \Delta {\rm Wr}/L$. Now} the writhe Wr depends on the molecular axis shape $\mathbf{r}(s)$ only, and not on the possible local torsion {\em inside} the polymer~\cite{Marko2015,Klenin2000}. {Thus for a regular plectoneme, $\Delta {\rm Wr}$ can only depend on $R_0$ and $\Omega_0$. So does $\Delta \tau_0$.}

Note also that $\Delta \tau_0$ is the {{\em Frenet-Serret}} torsion variation associated with the geometrical torsion of the molecular axis defined by the helicoidal curve $\br(s)$.

In the case of a ribbon, an additional {\em relative} torsion contribution, that we will denote by  $a(s)$, comes from the fact that the ribbon can twist around its molecular axis~\cite{Chouaieb2006,Moffatt1992}. More precisely, the ribbon ``generatrix'' $\mathbf{u}(s)$ can make a non-zero angle $\phi$ with respect to the normal $\mathbf{n}(s)$ in the Frenet-Serret frame, {sometimes called the ``register''~\cite{Chouaieb2006}. Then} $a(s)={\rm d}\phi/{\rm d}s$. {More details are given in Appendix~\ref{ribbon} and figure~\ref{ribbon:fig}}. In principle, the local torsion of the dsDNA is thus allowed to fluctuate, and the torsional energy reads
\begin{equation}
\frac{C}2 \int_{-L/2}^{L/2} {\rm d}s \, [\sigma \omega_0 + \Delta \tau_0 + a(s)]^2
\label{a:s}
\end{equation} 
We can minimize this free energy with the constraint $\int a(s) \, {\rm d}s = 0$ coming from the conservation of twist through the conservation of writhe at fixed plectoneme molecular axis shape $\mathbf{r}(s)$ {(see Appendix~\ref{ribbon})}. We are led to $a(s) \equiv 0$. 

\begin{figure}[ht]
\begin{center}
\includegraphics[height=4.5cm]{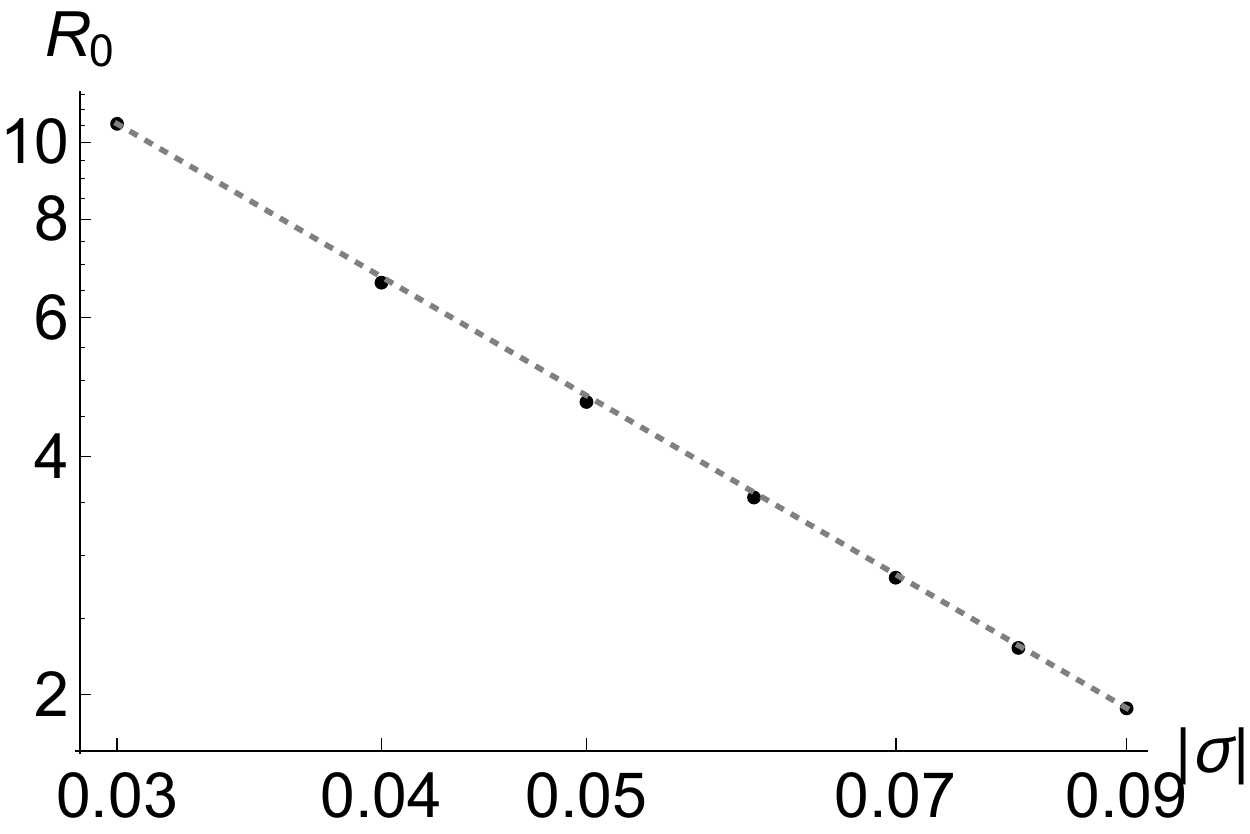}~~~\includegraphics[height=4.5cm]{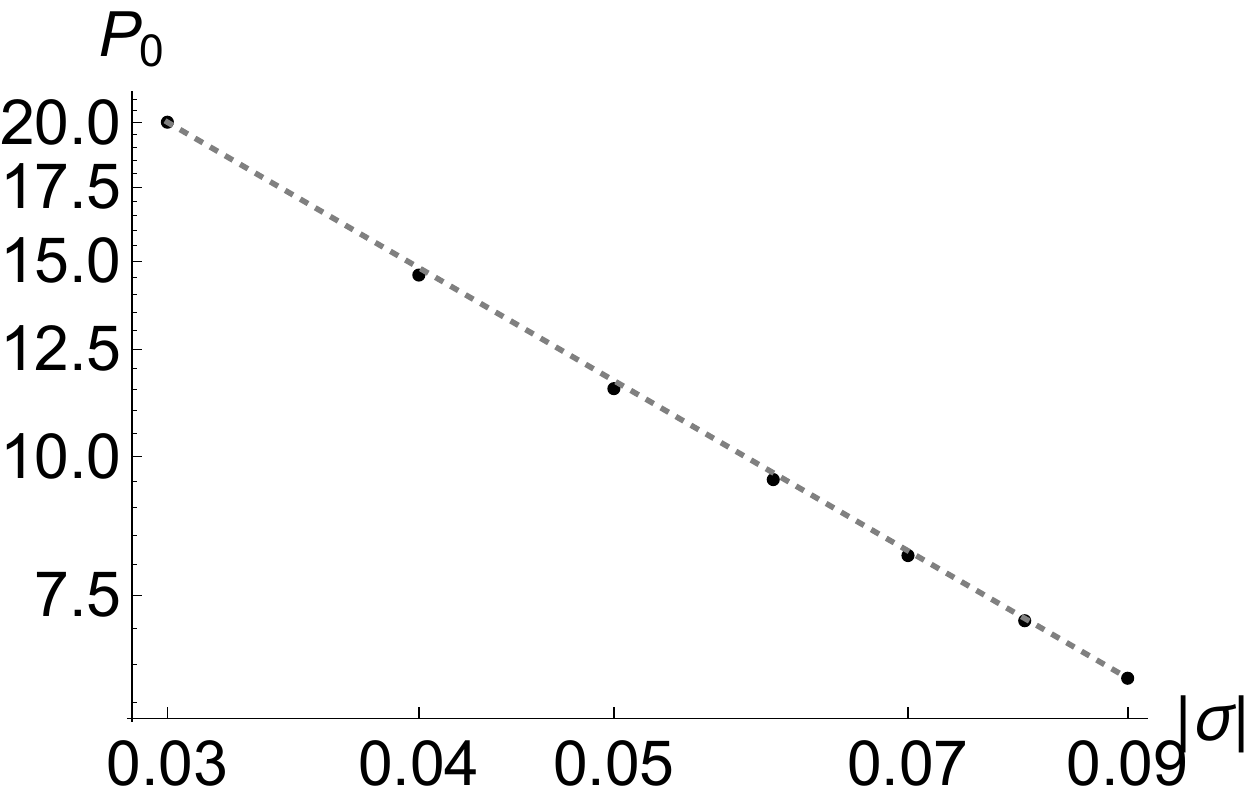}
 \caption{The plectoneme radius $R_0$ (left) and its pitch $P_0$ (right), both in nm, in function of the supercoiling density $\sigma$ for $C=110$~nm and $\omega_0 = 2\pi/p =  1.76$~rad/nm. The dotted lines are power laws with exponents $-1.55$ (left) and -1.05 (right), in agreement with Ref.~\cite{Marko2015} (exponents $-3/2$ and $-1$ respectively). Log-log coordinates. 
 \label{R0P0}}
 \end{center}
\end{figure}

Ignoring the thermal chain fluctuations (see the Discussion section below), the energy density follows 
\begin{equation}
f_0( R_0, \Omega_0) \equiv \frac{F}{L} = \frac{\ell_p}2 R_0^2 \,  \Omega_0^4 
+ \frac{C}2 \left[ \sigma \omega_0 + \Omega_0 
	\sqrt{1 - R_0^2 \Omega_0^2}  \right]^2 + \frac{K}{R_0^{2/3}}.
\label{order0}
\end{equation}
Minimization with respect to $R_0$ and $\Omega_0$ yields 
\begin{eqnarray}
\frac{\partial f_0}{\partial R_0} & = & \ell_p R_0 \,  \Omega_0^4 - C \left[ \sigma \omega_0 + \Omega_0 
	\sqrt{1 - R_0^2 \Omega_0^2}  \right] 
	\frac{R_0 \Omega_0^3}{\sqrt{1 - R_0^2 \Omega_0^2}} - \frac23 \frac{K}{R_0^{5/3}} = 0 \label{equil1} \\
 \frac{\partial f_0}{\partial \Omega_0} & = & 2 \ell_p R_0^2 \,  \Omega_0^3 +  C \left[ \sigma \omega_0 + \Omega_0 
	\sqrt{1 - R_0^2 \Omega_0^2}  \right]   
	\frac{1-2 R_0^2 \Omega_0^2}{\sqrt{1 - R_0^2 \Omega_0^2}}=0 \label{equil2} 
\end{eqnarray}
which can be solved numerically, leading to:
\begin{eqnarray}
R_0 &= &4.7~{\rm nm}~({\rm resp.}~5.3~{\rm nm})\\
\Omega_0 & = & 0.080~{\rm nm}^{-1}~({\rm resp.}~0.074~{\rm nm}^{-1})
\end{eqnarray}
with the above parameter values, notably $C=110$~nm (resp. 75~nm) and $\omega_0 = 2\pi/p{/\ell_0} =  1.76$~rad/nm ($p=10.5$~bp per ds-DNA helix turn {and $\ell_0=0.34$~nm, the ds-DNA rise per base-pair}). The supercoiling density is chosen as $\sigma=-0.05$, in absolute value well above the limiting threshold for plectoneme stability~\cite{Marko2015}. In biological DNA, $\sigma$ fluctuates around this typical value~\cite{Vologodskii2015}, thus it will be varied below. With this value, the plectoneme pitch is $P_0=12.4$~nm. Additional values of $R_0$ and $P_0$ are given in figure~\ref{R0P0}. The orders of magnitude of $R_0$ and $P_0$ are quite realistic from an experimental or numerical point of view (see, e.g., Refs.~\cite{Fathizadeh2014,Boles1990}). The observed scalings with $\sigma$ are consistent with the predictions of Marko in the limit $P_0 \gg R_0$ and large enough $|\sigma|$ ($|\sigma| \gg 10^{-3}$ in ds-DNA; see Ref.~\cite{Marko2015}, section 4.1.4).

\subsection{Pinched plectoneme ($\lambda \neq 0$)~-- small-pinching limit}

Figure~\ref{R0P0} shows that the plectoneme radius $R_0$ is typically larger than the capture radius $R_c$ set by the protein or protein complex size, on the nanometer range. To compute the elastic energy required to pinch the plectoneme at $s=0$ we impose a small force $\lambda \equiv \varepsilon \lambda_1$, where $\varepsilon$ is the unique small parameter of the problem. 

In the spirit of the perturbation calculation proposed by Marko and Siggia~\cite{Marko1995}, though following a different route because our goals are different, we denote by $\mathbf{r}_0(s)$ the original position of the point of curvilinear abscissa $s$, and by $\mathbf{r}(s)$ its new, perturbed position~\cite{Marko1995}:
\begin{equation}
\mathbf{r}(s) = \mathbf{r}_0(s) + \varepsilon \, \mathbf{r}_1(s) + \mathcal{O}(\varepsilon^2)
\end{equation}
at order 1 in $\varepsilon$. We adopt the equivalent representation as proposed in equation~\eqref{main:eq}. It amounts to use the cylindrical coordinate system $(R(s),\theta(s)=\Omega(s)s,z(s))$, where we anticipate $R(s) = R_0 + \varepsilon r_1(s) + \mathcal{O}(\varepsilon^2) $ and $\Omega(s) = \Omega_0 + \varepsilon \omega_1(s) + \mathcal{O}(\varepsilon^2)$ (or alternatively $\theta(s) \equiv \Omega(s) \, s=  \Omega_0 \, s + \varepsilon \theta_1(s) + \mathcal{O}(\varepsilon^2)$ \footnote{With $\theta_1(s) \equiv \omega_1(s) \, s$.}) at order 1. We shall work at the lowest relevant order in the small parameter $\varepsilon$ (order 2 in practice, see below). 

\begin{figure}[ht]
\begin{center}
\includegraphics[height=12cm, angle=-90]{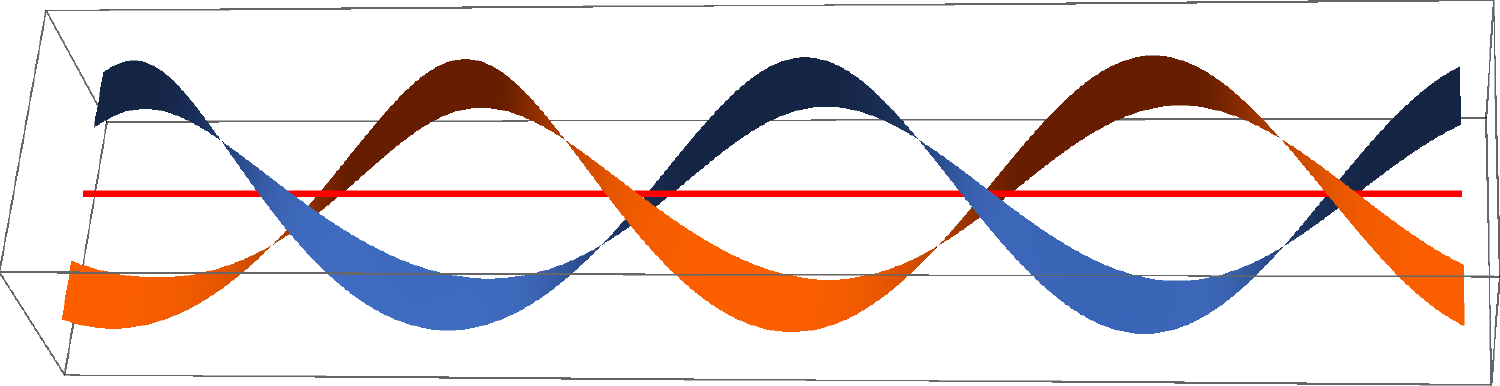} \\ \vspace{5mm}
\includegraphics[height=14.7cm, angle=-90]{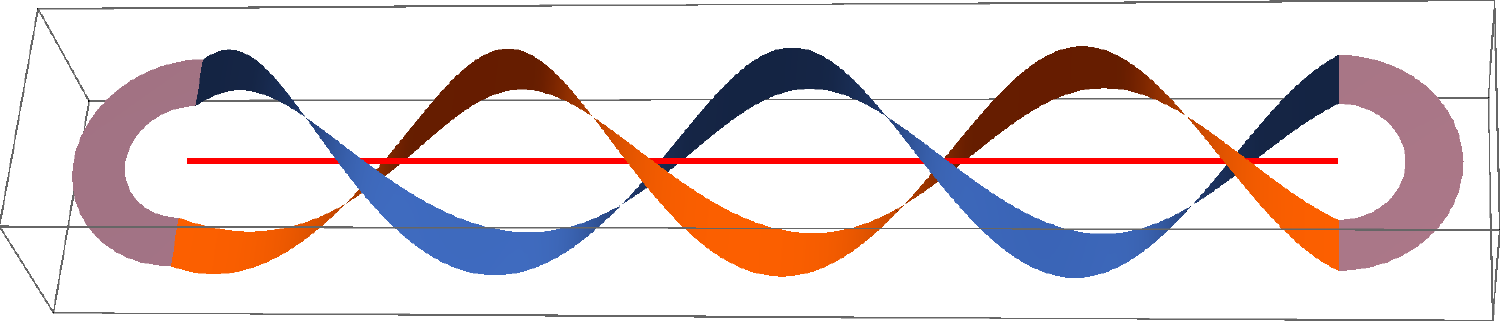}
 \caption{Top: The ``false'' ribbon plectoneme used for the calculations, before being pinched at the origin. Its generatrix $\mathbf{u}(s)$ is always perpendicular to the $z$-axis, displayed in red (compare to figure~\ref{fig0}). Bottom: If the ribbon is closed at its extremities by two half-annuli (represented in gray), its linking number Lk is exactly zero. 
 \label{fig1}}
 \end{center}
\end{figure}

One of the difficulties of the calculation is to deal with the possible local variations $a(s) = \varepsilon a_1(s) + \mathcal{O}(\varepsilon^2)$ of the relative over-torsion of the double strand as in equation~\eqref{a:s}.  This is potentially an additional way to relax the elastic constraint \emph{inside} the double strands~\cite{Chouaieb2006}.  For sake of conveniance, we calculate these variations with respect to a ``false ribbon'' (see figure~\ref{fig1}) where $\mathbf{u}(s) = \mathbf{u}_0(s)\equiv \mathbf{e}_r$~\footnote{Here $\mathbf{e}_r$ is the radial vector in cylindrical coordinates; the cylindrical frame is denoted by $(\mathbf{e}_r,\mathbf{e}_\theta,\mathbf{e}_z )$.}. The advantage of this choice is that $\mathbf{u}(s)$ then coincides with the normal $\mathbf{n}(s)$ thus the ribbon torsion and the molecular axis torsion coincide~\cite{Chouaieb2006}. To sum up, this false ribbon represents an imaginary molecule \emph{without} intrinsic supercoiling.  In the original ribbon \emph{with} intrinsic supercoiling $\sigma \omega_0 \neq 0$, only {the \emph{variations} of of the ribbon torsion coincide with the molecular axis torsion (Appendix~\ref{ribbon})}. We shall come back to this original ribbon at the end of the calculation. 

Note that, in addition, if the false ribbon is closed at its extremities by two half-annuli as in figure~\ref{fig1} (bottom), its linking number Lk is exactly zero whatever the length $L$ if there is an even number of crossings. Hence ${\rm Wr}=-{\rm Tw}$.

\medskip

\noindent \emph{Nota}: After pinching, in addition to $R(s) = R_0 + \varepsilon r_1(s) +\mathcal{O}(\varepsilon^2)$ and $\Omega(s) = \Omega_0 + \varepsilon \omega_1(s)+\mathcal{O}(\varepsilon^2)$, we also have $\mathbf{u}(s) = 
 \mathbf{e}_{r,0} + \varepsilon \mathbf{u}_1(s) +\mathcal{O}(\varepsilon^2)$~\footnote{The initial cylindrical frame (before pinching) is now denoted by $(\mathbf{e}_{r,0} \equiv \mathbf{u}_0,\mathbf{e}_{\theta,0},\mathbf{e}_z )$.}. This correction takes the torsion variation $a_1(s)$ into account. It also ensures that $\mathbf{u}(s)$ remains orthogonal to $\mathbf{t}(s)$. Note that $\mathbf{u}_1(s)$ is not necessarily a unitary vector. By contrast, both $\mathbf{u}(s)$ and $\mathbf{e}_{r,0}$ are unitary, thus $\mathbf{e}_{r,0} \cdot \mathbf{u}_1 = \mathcal{O}(\varepsilon)$.

\medskip

Far from the origin, we expect that $R(s) \rightarrow R_0$ so that the elastic energy remains finite, as well as $\theta(s)-\Omega_0 s \rightarrow \pm \Delta \theta $ when $s \rightarrow \pm \infty$. Here $\Delta \theta = \varepsilon \Delta \theta_1 + \mathcal{O}(\varepsilon^2)$ is an {\em a priori} allowed global rotation (around the $z$-axis) of the extremities due to the pinching. Indeed, the plectoneme extremities are free to rotate arount $(Oz)$ to relax at least partially the pinching constraint. We also expect $a_1(s) \rightarrow 0$ far from the origin, where the polymer is not perturbed. 

\subsubsection{{Order $\varepsilon$}}

We first compute the bending and twisting contributions to the free energy in equation~\eqref{F:gen}. 

\medskip
\noindent \emph{a}.~\underline{Bending:} We need to expand at order 1 the position $\mathbf{r}(s)$ of equation~\eqref{r:of:s}. We first write $\mathbf{r}(s) = R(s) \, \mathbf{e}_r + z(s) \,  \mathbf{e}_z$ (here we use the {\em new} cylindrical frame)
 and we calculate the squared curvature
\begin{eqnarray}
\gamma^2(s) & \equiv & \left\| \frac{{\rm d} \mathbf{t}}{{\rm d}s}  \right\|^2 = \left\| \frac{{\rm d}^2 \mathbf{r}}{{\rm d}s^2}  \right\|^2 \\
& = & \left[R (\theta')^2 - R''\right]^2 + \left[2 R' \theta' + R \theta''\right]^2  + (z'')^2 \\
& = & \left[(R_0+ \varepsilon \, r_1)  (\Omega_0 + \varepsilon \, \theta_1')^2 - \varepsilon \, r_1'' \right]^2 
 \nonumber \\
& & +  \left[ 2  \varepsilon \,r_1' (\Omega_0 + \varepsilon \, \theta_1') + (R_0+ \varepsilon \, r_1) \varepsilon \theta_1''\right]^2 + (z'')^2 + \mathcal{O}(\varepsilon^2).
\label{C2}
\end{eqnarray}
One checks that 
\begin{equation}
z''(s)=-\varepsilon\, \frac{R_0 \Omega_0}{\sqrt{1-R_0^2 \Omega_0^2}}[R_0 \theta_1''(s)+\Omega_0 r_1'(s)] + \mathcal{O}(\varepsilon^2). 
\end{equation}
Thus $(z'')^2$ is of order $\varepsilon^2$, as well as the second term of the r.h.s. of equation~\eqref{C2}. At order 0, $\gamma_0^2=R_0^2 \Omega_0^4$, as expected from equation~\eqref{gamma0}.  The order-1 corrections to the bending energy density $(\ell_p/2) \gamma_0^2$ in equation~\eqref{order0} are thus $\ell_p \gamma_0 \gamma_1(s)$ with 
\begin{equation}
\gamma_1(s) = \Omega_0^2 r_1(s) + 2 R_0 \Omega_0 \theta_1'(s) - r_1''(s).
\end{equation}

\medskip

\noindent \emph{b}.~\underline{Twisting:} If $a(s)=0$, we recall that the geometrical torsion can be calculated through the Frenet-Serret formula $\tau = - {\rm d} \mathbf{b} / {\rm d}s \cdot \mathbf{n}$, where $\mathbf{b}=\mathbf{t} \times \mathbf{n}$ is the binormal. After calculation, the order-1 corrections to the torsion are  
\begin{equation}
\tau_1(s) =   \frac1{\sqrt{1-R_0^2 \Omega_0^2}} \left[
- {R_0 \Omega_0^3 } r_1(s) - \frac{2-R_0^2 \Omega_0^2}{R_0 \Omega_0} r_1''(s)
+ (1 - 2 R_0^2 \Omega_0^2 ) \theta_1'(s) 
- \frac1{\Omega_0^2} \theta_1^{(3)} \right]
+a_1(s).
\label{tau1:s}
\end{equation}
Here we have introduced the additional order-$\varepsilon$ correction, $a_1(s)$, due to the internal over-torsion of the double strand. This result again expresses the fact that the torsion variation is the sum of the ({\em Frenet-Serret}) geometrical torsion variation and of the {\em relative} torsion variation $a_1(s)$~\cite{Chouaieb2006,Moffatt1992} {(see also Appendix~\ref{ribbon})}. The order-$\varepsilon$ correction to the torsional energy density  in equation~\eqref{order0} is thus $C \Delta \tau_0 \tau_1(s)$, i.e., coming back to the original ribbon \emph{with} intrinsic supercoiling $\sigma \omega_0$, $C \left[ \sigma \omega_0 + \Omega_0 
	\sqrt{1 - R_0^2 \Omega_0^2}  \right] \tau_1(s)$.

\medskip

\noindent \emph{c}.~\underline{Minimization of the free energy:} The free energy $F$ of equation~\eqref{F:gen} is a functional of $r_1(s)$, $\theta_1(s)=\omega_1(s) s$, $\psi_1(s)$, and their derivatives with respect to $s$. When $s \rightarrow \pm \infty$, these quantities are constrained by: $r_1(s) \rightarrow 0$ and $a_1(s) \rightarrow 0$ so that the elastic energy remains finite, as well as $\theta'_1(s) \rightarrow 0$ (or $\theta_1(s) \rightarrow \pm \Delta \theta_1$). The order-1 corrections to the order 0 in equation~\eqref{order0} are thus 

\begin{eqnarray}
F_1 & = & \int_{-L/2}^{L/2} f_1(s) \; {\rm d} s \quad \mbox{with} \\
f_1(s) & = &   \ell_p  R_0 \Omega_0^2 \, \gamma_1(s) + C \left[ \sigma \omega_0 + \Omega_0 
	\sqrt{1 - R_0^2 \Omega_0^2}  \right] \tau_1(s) - \frac23 \frac{K}{R_0^{5/3}} \, r_1(s)
\end{eqnarray}
First one integrates the term of $f_1$ depending on $a_1(s)$. The integral vanishes, $\int_{-L/2}^{L/2} a_1(s) {\rm d}s=0$,  because the total twist variation $\Delta {\rm Tw}= (1/2\pi) \int_{-L/2}^{L/2} a(s) {\rm d}s$ is fixed through the conservation of the linking number Lk at given polymer shape $\mathbf{r}(s)$, as explained above {and in Appendix~\ref{ribbon}}.

By contrast  $\theta_1$ does not need to vanish at large $L$ and the integral of the terms linear in $\theta_1'(s)$ is
\begin{equation}
2 \Delta \theta_1 \left\{ 2 \ell_p   R_0^2 \Omega_0^3   + C \left[ \sigma \omega_0 + \Omega_0 
	\sqrt{1 - R_0^2 \Omega_0^2}  \right] \frac{1 - 2 R_0^2 \Omega_0^2 }{\sqrt{1-R_0^2 \Omega_0^2}} \right\}
	\equiv 2 \Delta \theta_1 \, A_\theta.
\label{A_s}	
\end{equation}
It vanishes owing to equation~\eqref{equil2}. Finally, the minimization of $F_1$ with respect to $r_1(s)$ reads:
\begin{equation}
0 = \frac1{\varepsilon} \frac{\delta F_1}{\delta r_1(s)}  =   \ell_p  R_0 \Omega_0^4 - C \left[ \sigma \omega_0 + \Omega_0 
\sqrt{1 - R_0^2 \Omega_0^2}  \right]  \frac{R_0 \Omega_0^3 }{\sqrt{1-R_0^2 \Omega_0^2}} - \frac23 \frac{K}{R_0^{5/3}} \equiv A_r.
\label{A_theta}
\end{equation}
This last equation is automatically satisfied through equation~\eqref{equil1}. Note that the term $\lambda [R(0] - R_c]$ in equation~\eqref{F:gen} does not contribute at this order because $\lambda$ is of order one and the correction $\varepsilon r_1(0)$ to $R_0$ is also of order 1. This term will start contributing at the order 2 below. 

To conclude, the order 1 in $\varepsilon$ does not bring additional information as compared to order 0. 

\medskip

\noindent \emph{Nota}: In $f_1(s)$, the prefactors of $r_1(s)$ and $\theta'_1(s)$, that we have respectively denoted by $A_r$ and $A_\theta$, vanish, as displayed in equations~\eqref{A_s} and \eqref{A_theta}. This could be anticipated because in the special case where 
$r_1(s)$ and $\theta'_1(s)$ are respectively replaced by the constant functions $\delta R_0$ and $\delta \Omega_0$ in $f_1$, the total free energy $F_0$ becomes $F_0 + \delta F_0$ with $\delta F_0 = A_r \, L \, \delta R_0 + A_\theta \, L \, \delta \Omega_0$. Thus $A_\theta = (1/L) \partial F_0 / \partial \Omega_0$ and $A_r = (1/L) \partial F_0 / \partial R_0$ and they vanish automatically. 

\medskip

\subsubsection{{Order $\varepsilon^2$}}

As it could have been anticipated, order $\varepsilon$ is trivial. But dealing with it brought some interesting insight into the plectoneme elasticity that will be useful below. We are thus led to go to order $\varepsilon^2$ to compute the linear elastic response to pinching:
\begin{eqnarray}
R(s) & = & R_0 + \varepsilon r_1(s) +\varepsilon^2 r_2(s) + \mathcal{O}(\varepsilon^3), \\
\theta(s) & = & \Omega_0 s + \varepsilon \theta_1(s)+ \varepsilon^2 \theta_2(s)+\mathcal{O}(\varepsilon^3), \\
a(s) & = & \varepsilon a_1(s) + \varepsilon^2 a_2(s) +\mathcal{O}(\varepsilon^3),
\end{eqnarray}
and so on. Order-2 terms in the free energy will be either the products of order-1 terms quadratic in the functions $r_1(s)$, $\theta_1(s)$, $a_1(s)$ and their derivatives; or the products of order-0 and order-2 terms proportional to $r_2(s)$, $\theta_2(s)$, $a_2(s)$ and their derivatives. The latter will not contribute to the free energy $F_2$, in the same way as the order-1 terms above: the calculations would be exactly the same as above, just replacing the functions $r_1(s)$, $\theta_1(s)$, or $a_1(s)$ by $\varepsilon r_2(s)$, $\varepsilon \theta_2(s)$, $\varepsilon a_2(s)$. 

The quadratic parts of the order-2 corrections associated with bending, twisting and confinement are
denoted by $Q_b(s)$, $Q_t(s)$ and $Q_c(s)$ respectively. Although $Q_b(s)$ and $Q_c(s)$ are relatively straightforward, the calculation of $Q_t(s)$ is more tricky. {Their full, somewhat lengthy, calculation is given in Appendix~\ref{quads}.}


Given that only $\theta'_1$ and its derivatives appear in the free energy (not $\theta_1$ itself), we switch to the variable set $r_1(s)$, $\theta'_1(s)$ and $a_1(s)$. As explained above we will get the linear, elastic response to pinching by minimizing the functional quadratic form
\begin{equation}
F_2[r_1(s),\theta'_1(s),a_1(s)] = \int [ Q_b(s) + Q_t(s)+Q_c(s)] {\rm d}s 
\end{equation}
with respect to the fields $u(s)\equiv r_1(s)$, $v(s) \equiv \theta'_1(s)$ and $w(s) \equiv a_1(s)$.  We switch to the Fourier representation by using Parseval's theorem:
\begin{equation}
F_2[\hat u, \hat v, \hat w] = \int \frac{{\rm d} q}{2 \pi} \, \frac12  \, ^t\bar{U}(q) M(q)  U(q)
\end{equation}
where the vector $U(q)$ has coordinates $(\hat u(q), \hat v(q), \hat w(q))$ and the expression of the $3 \times 3$ matrix $M(q)$ is given in Appendix~\ref{full:M}.

In equation~\eqref{F:gen}, we had already introduced the pinching constraint on $R(0)$ through the Lagrange multiplier $\lambda = \lambda_1 \, \varepsilon + \mathcal{O}(\varepsilon^2)$. We now have to deal with the additional constraint $\int_{-L/2}^{L/2} a(s) {\rm d}s=0$ on $a(s)$ {due to the fact that the molecule is closed at its extremities}, as explained above. We enforce it through a second Lagrange multiplier denoted by $\mu=\mu_1\, \varepsilon + \mathcal{O}(\varepsilon^2)$. In the Fourier space, the total functional to be minimized at order 2 in $\varepsilon$ is thus 
\begin{equation}
G_2[\hat u, \hat v, \hat w] = F_2[\hat u, \hat v, \hat w] -\int \frac{{\rm d} q}{2 \pi} \, \left[ \lambda_1 \,\hat u(q) + \mu_1 \, 2 \pi \delta(q) \,\hat{w}(q) \right]
\end{equation}
where $\delta(q)$ is Dirac's distribution in the Fourier space. Minimization of $G_2$ yields
\begin{equation}
M(q)  U(q) = \left( 
\begin{array}{c}
\lambda_1 \\ 0 \\ {2 \pi} \mu_1 \delta(q)
\end{array}
  \right)
\end{equation}
from which the fields $\hat u(q)$, $\hat v(q)$ and $\hat w(q)$ can now be simply inferred by inversion of the matrix $M(q)$.

By inverse Fourier transform, we can then express the solutions $r_1(s)$, $\theta'_1(s)$ and $a_1(s)$ in function of both $\lambda$ and $\mu$. Notably 
$a_1(s)= \lambda_1 \mathcal{F}^{-1}\left[ M^{-1}_{31}(q)\right] + \mu_1 M^{-1}_{33} (0)$ ($\mathcal{F}^{-1}$ is the inverse Fourier transform).

\begin{figure}[t!]
\begin{center}
\includegraphics[height=4.5cm]{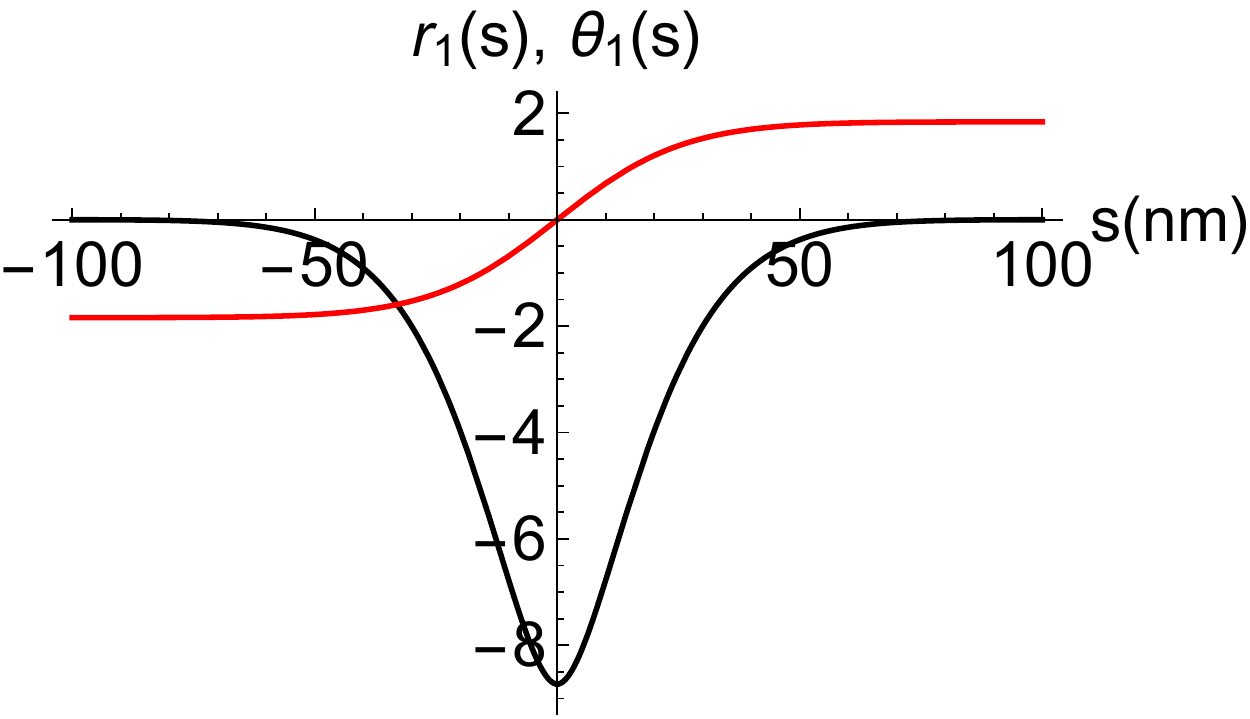} \includegraphics[height=4.5cm]{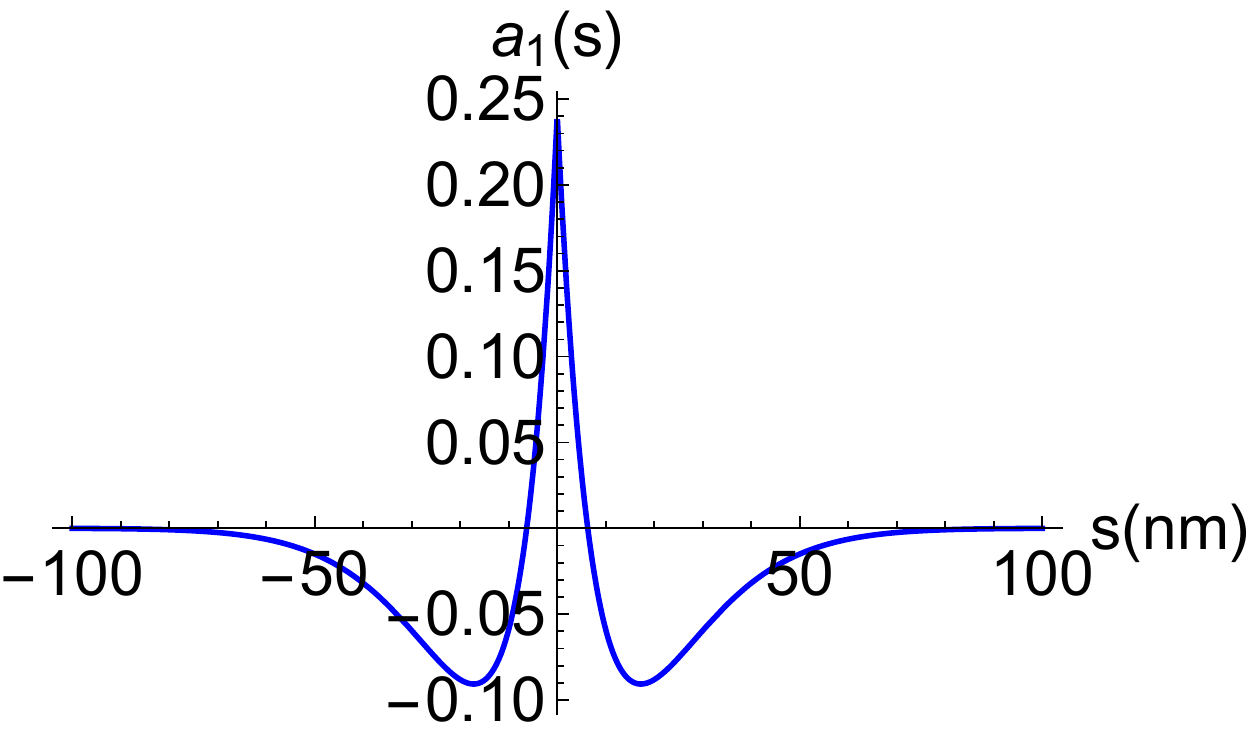} 
 \caption{Analytical solution of the linear response of the plectoneme to pinching at $s=0$ in the case $\lambda_1=-1$. The following parameter values have been used: $\ell_p=50$~nm, $C=110$~nm, $p=10.5$~bps/turn, $c=1$ and $\sigma=-0.05$ (see text for the definitions). Left: $r_1(s)$ (black, in nm) and $\theta_1(s)$ (red, in rad). Right: $a_1(s)$, in rad/nm.
 \label{solutions}}
 \end{center}
\end{figure}

The value of $\mu_1$ is then set by imposing effectively that $\int_{-L/2}^{L/2} a_1(s) {\rm d}s=0$. It follows that $\mu_1 = \mathcal{O}(1/L)$ and eventually that at large $L$, 
\begin{eqnarray}
r_1(s) & = & \lambda_1 \mathcal{F}^{-1} \left[ M^{-1}_{11}(q) \right] \\
\theta'_1(s) & = & \lambda_1 \mathcal{F}^{-1} \left[ M^{-1}_{21}(q) \right] \\
a_1(s) & = & \lambda_1 \mathcal{F}^{-1} \left[ M^{-1}_{31}(q) \right] \label{a1}
\end{eqnarray}
These solutions are linear combinations of the exponentials $e^{i \, \alpha_k s}$ where the $\alpha_k$'s are the 6 complex roots of $\det M(q)$. An example is displayed in figure~\ref{solutions}. The deformation range is the inverse of the smallest eigenvalue imaginary part, and is on the order of 30~nm in this case. 

\begin{figure}[ht]
\begin{center}
\includegraphics[height=5cm]{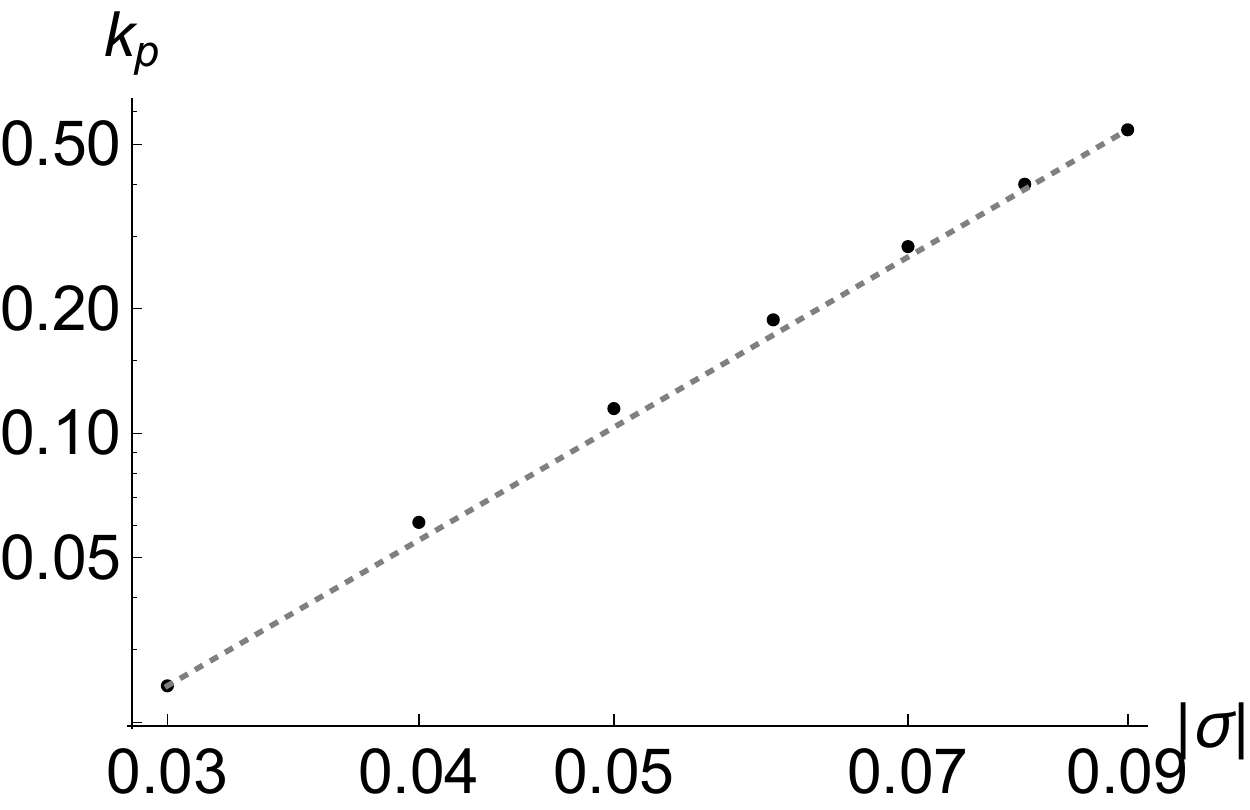} \quad \includegraphics[height=5cm]{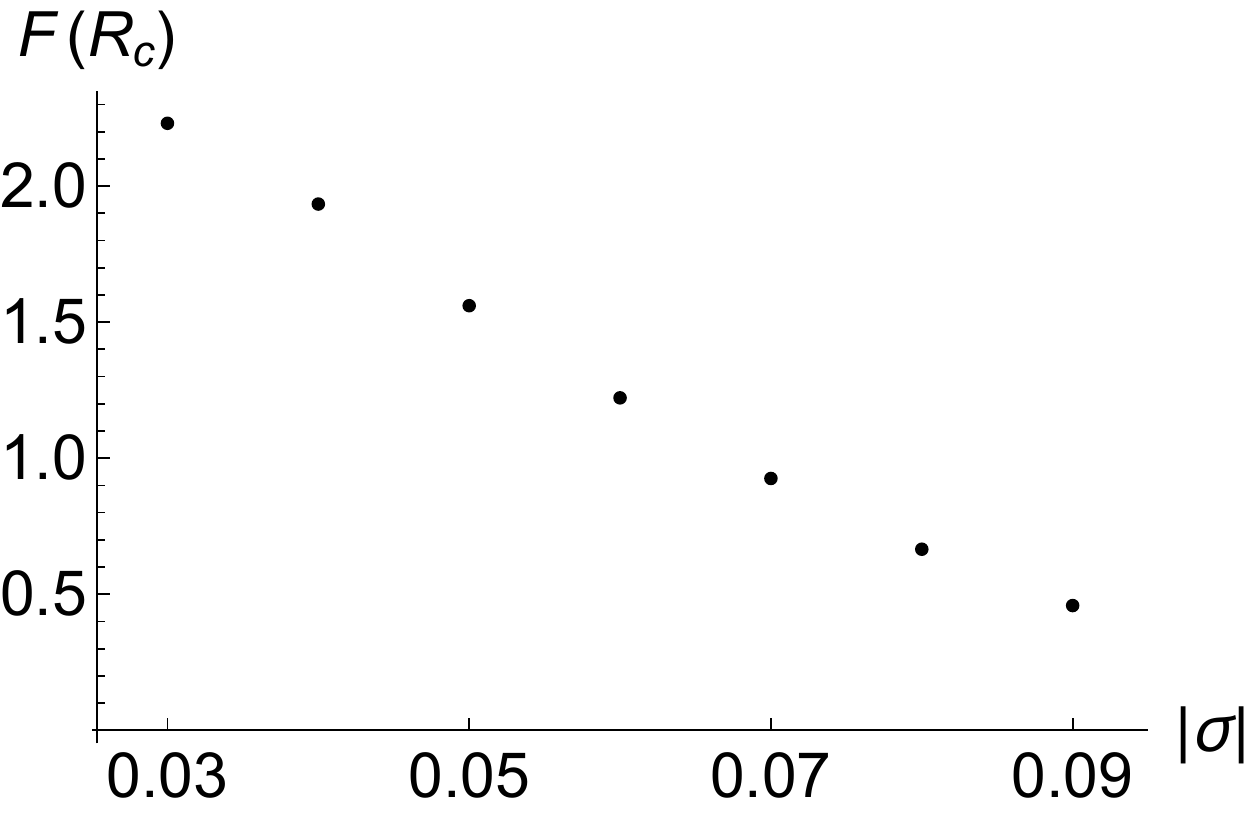} 
 \caption{Left: Spring constant $k_p$ characterizing the linear response of the plectoneme to pinching in function of the  supercoiling density $\sigma$, as given in equation~\eqref{k:eq}, in units of $k_{\rm B}T/{\rm nm}^2$. Log-log coordinates; The dotted lines is a fitted power law with exponent 2.82. Right: Pinching energy against $\sigma$ and in units of $k_{\rm B}T$, assuming a capture radius $R_c=1$~nm. Same parameters as in figure~\ref{solutions}.
 \label{k}}
 \end{center}
\end{figure}

Once $r_1(s)$ is known, the linear response of the plectoneme to the pinching force $\lambda$ is characterized by the spring constant $k_p$ such that 
\begin{equation}
k_p^{-1} = \lim_{\varepsilon \rightarrow 0} \frac{R(s)-R_0}{\lambda} = \frac{r_1(0)}{\lambda_1} =  \mathcal{F}^{-1}  \left[ M^{-1}_{11}(q) \right] \Big|_{s=0} = \int_{-\infty}^\infty \frac{{\rm d}q}{2 \pi} \, M^{-1}_{11}(q) 
\label{k:eq}
\end{equation}
The spring constant $k_p$ is plotted in function of the supercoiling density $\sigma$ in figure~\ref{k} (left). With the chosen parameters, the data are well fitted by a power law with exponent 2.82.
 
From the value of $k_p$, one computes the work of $\lambda$ needed to pinch the plectoneme down to the capture radius $R_c$:
\begin{equation}
F(R_c) = k_p (R_0-R_c)^2 .
\label{F:eq}
\end{equation}
This expression takes into account the fact that the same work is needed for both strands. This is the free-energy barrier to be overcome in order to form the enhancer-promoter-protein(s) synapse.  It is also plotted in function of  $\sigma$ in figure~\ref{k} (right) for $R_c=1$~nm.

\subsection{Role of torsion/twist}

Since $\mu_1=0$ in the large $L$ limit, as established in the previous subsection, it appears that $\delta  G_2/ \delta a_1(s) = C \tau_1(s) \equiv 0$. Coming back to equation~\eqref{tau1:s}, this means that the relative torsion $a_1$ exactly compensates the Frenet-Serret, geometrical torsion variation, as it was in fact anticipated below equation~\eqref{a:s}. Without {\em a priori} considering it as being established, we confirmed this results through the full calculation taking all degrees of freedom into account on an equal footing. Relaxing the pinching constraint through $a(s)$ thus lowers the system energy as compared to a plectoneme made of two ideal polymers without internal twist degrees of freedom $a(s)$.

\subsection{{Role of electrostatics}}

We have assumed so far that the electrostatic short-range contribution of the repulsive energy is negligible in the free energy \eqref{F:gen}. 
This is justified when the plectoneme radius remains large as compared to the Debye screening length, $\lambda_D \approx 0.8$~nm at physiological conditions. This is true when the plectoneme is unperturbed because $R_0$ is larger than $\lambda_D$ for the regime of parameters explored in the work. However the inter-strand distance $2R(s)$ can become on the order of  $\lambda_D$ when the plectoneme is pinched. To quantify this effect, we use the form of the electrostatic energy $U(R,P)$ per persistence length between two intertwined polyelectrolytes proposed in Ref.~\cite{Brahmachari2017} (see their equation~(6), as well as Ref.~\cite{Ubbink1999}). For example, we find that if $|\sigma|=0.05$, $R_0\simeq 4.7$~nm and $P_0 \simeq 11.6$~nm, then $U(R,P)\simeq 0.2~k_{\rm B}T$ per persistence length, which is indeed negligible. If $|\sigma|=0.07$ (then $R_0\simeq 2.8$~nm and $P_0 \simeq 8.1$~nm) and $U(R,P)$ grows to $\simeq 1.5~k_{\rm B}T$ per persistence length, and up to $13.6~k_{\rm B}T$ per persistence length if $|\sigma|=0.09$ (then $R_0\simeq 1.9$~nm and $P_0 \simeq 6.3$~nm).

To go further, we must estimate the additional work $W_{\rm elec}$ required to pinch the two helices when taking into account electrostatic repulsion, as follows. We suppose that the equation~(6) giving $U(R,P)$ in Ref.~\cite{Brahmachari2017} for an un-pinched helix remains valid for a pinched one by just replacing the radius of the helix $R_0$ by its local value $R(s)$ inferred from the expression of $a_1(s)$, because $R(s)$ varies slowly with $s$. 
The value of $P$ also depends on $s$ as follows. Owing to equation~\eqref{Om0}, we set
\begin{equation}
P(s) \simeq \sqrt{\frac1{\Omega^2(s)}-R^2(s)},
\end{equation}
where $\Omega(s) \equiv \theta(s)/s \simeq \Omega_0 + \varepsilon \theta_1(s)/s$ at order $\varepsilon$. Injecting $R(s)$ and $P(s)$ in  
the expression of $U(R,P)$, where we set the effective charge $\nu\approx6.2$~nm$^{-1}$ at monovalent salt concentration 0.1~M~\cite{Brahmachari2017}, it follows that 
\begin{equation}
W_{\rm elec} \simeq \int_{-\ell_p}^{\ell_p} \frac{U(R(s),P(s))-U(R_0,P_0)}{\ell_p} \, {\rm d}s.
\end{equation} 
As explained above, the value $R_c=1$~nm was willingly chosen smaller than the real enhancer-promoter-protein synapse size because we aimed at providing an upper bound of the energy required to bring the two opposed ds-DNA strands closer. A more probable synapse size is in the 5~nm range~\cite{Liu2016,Zhang2011}, which gives $R_c = 2.5$~nm. For example, if $|\sigma|=0.05$, $R_0\simeq 4.7$~nm and $P_0 \simeq 11.6$~nm, which are typical values for a real plectoneme, then we obtain that $W_{\rm elec} \simeq 0.95~k_{\rm B}T$, below the thermal energy $k_{\rm B}T$. For $|\sigma|\leq 0.07$,  $W_{\rm elec}$ remains smaller than $k_{\rm B}T$. Note that above this value of $|\sigma|$, $R_0$ becomes smaller than $R_c=2.5$~nm. We cannot speak of ``pinching'' anymore. 

By contrast, choosing a smaller value of $R_c$ leads to estimates of $W_{\rm elec}$ above the $k_{\rm B}T$ range. For instance, if $R_c=2$~nm (resp. 1.5~nm) and  $|\sigma|=0.05$, $R_0\simeq 4.7$~nm and $P_0 \simeq 11.6$~nm, then $W_{\rm elec} \simeq~3.0~k_{\rm B}T$ (resp. $10.0~k_{\rm B}T$). An energy barrier larger than the thermal energy arises when the synapse is small. This suggests that the electrostatic contribution at short range cannot be ignored at very short capture radii and it ought to be added to our energy in equation~\eqref{F:gen}, even though it itself relies on some assumptions and approximations~\cite{Ubbink1999}. In particular, the electrostatic energy depends on the square of the effective linear charge density $\nu$, which takes into account counterions condensation. This issue is still under debate~\cite{Brunet2015} and the chosen value of $\nu$ is possibly overestimated, as well as the electrostatic repulsion. In addition, when both ds-DNA are very close, strong correlations between the counter-ion clouds occurs, which is not taken into account in Refs.~\cite{Brahmachari2017,Ubbink1999}. The protein-complex surface charge density as well as the counterions finite size might also play a role.

\section{Discussion and conclusion}

Taking into account all polymer internal and external degrees of freedom, we have been able to compute analytically the elastic response of a plectoneme to pinching at the origin. This was done in the approximation where thermal fluctuations are only partially taken into account through the effective repulsion between the two ribbons or the effective elastic parameters which themselves have an entropic contribution coming from the solvent and the complex atomic structure of the involved molecules. Going beyond this approximation and fully taking into account thermal fluctuations requires to appeal to Gaussian path integrals~\cite{Burkhardt1995} on the plectoneme shape (characterized by $U(q)$ in the Fourier space). At the quadratic level considered here, in the case where the force $\lambda$ is imposed to the system (i.e., {\em not} the plectoneme radius $R(0)$ at the origin), the quadratic form $M(q)$ does not depend on the intensity $\lambda$ of the force. Consequently, the free-energy contribution of fluctuations, $F_{\rm fluct.} = \frac12 k_{\rm B} T \int \log \det  M(q) \, {\rm d}q$, does not depend on $\lambda$ either and its calculation does not bring any additional information. 

An alternative way of tackling the problem would be to switch to a different statistical ensemble and  to impose $R(0)$ instead of $\lambda$. However, in the passive synapse assembly mechanism considered here, the plectoneme fluctuates until it can be captured by the transcription factor, and there is no reason to chose the statistical ensemble where $R(0)$ is fixed.

To go further, it would be necessary to fully take into account contacts between opposed strands {(or even more realistically, electrostatic interactions)} in the Hamiltonian $F_2$ and to apply tools from classical field theory to go beyond the effective repulsive interaction $\propto 1/R^{2/3}$ introduced in equation~\eqref{F:gen}. This is out of the scope of the present work. However, it is reasonable to expect that thermal fluctuations will not play a crucial role, beyond this entropic repulsion, because DNA is rather rigid at the sub-persistence length-scale under interest here. {More precisely, even though the plectoneme itself can be significantly modified by thermal fluctuations because the energies at play are on the $k_{\rm B}T$ range (see below), each dsDNA strand composing the plectoneme is only slightly bent: the local radius variation $\delta R = R_0-R_c$, of a few nanometers, is attained by bending each dsDNA on a distance $\delta s \simeq 40$~nm (figure~\ref{solutions}). The typical dsDNA bending angle $\delta R/\delta s\sim 0.1$~rad, is small. As for electrostatic interactions, we have seen that they might play a role for small transcription factor complexes, i.e. small capture radii below 2~nm, or high supercoling densities $|\sigma| > 0.07$.}

In figure~\ref{k} (right), the pinching energy $F(R_c)$ is plotted against the supercoiling density $\sigma$ for a capture radius $R_c=1$~nm. This value of $R_c$ is probably underestimated in the context of enhancer-promoter-protein synapse as discussed in the introduction because protein complexes involved in genetic machineries are generally larger. However, it sets an upper bound of the energy required to bring the two opposed ds-DNA strands closer: whatever the value of $\sigma$ in the biologically relevant range, $F(R_c)$ never exceeds $2\, k_{\rm B}T$. This proves that pinching can be achieved through thermal fluctuations in a short time and the ensuing energy barrier cannot explain the 1 to 10~ms time-scale observed in experiments~\cite{Bussiek2002} or simulations~\cite{Huang2001} for DNA minicircles as short as few thousand bp. 

If pinching is fast, then slithering is the slow step limiting the synapse formation (see figure~\ref{Barde}). The role of supercoiling is then simply to limit the accessible volume in the phase space, or differently said to decrease the translational entropy in the synapse open state. Supercoiling increases the equilibrium probability of the closed state with respect to the open one~\cite{Vologodskii1996,Huang2001}. From a kinetic point of view, slithering dynamics being governed by diffusion, the whole process is diffusion-limited. Plectoneme slithering has been studied in the DNA case by Marko and Siggia~\cite{Marko1995} through an analogy with polymer reptation~\cite{DeGennes1971}. If ignoring hydrodynamic interactions, the diffusion-slithering time $\tau_{\rm sl.}$ needed to bring the two sites at proximity on opposed plectoneme strands (as in figure~\ref{Barde}(b)) scales like $\eta L^3/k_{\rm B}T$ (up to some numerical prefactors), where $\eta$ is the solvent viscosity and $L$ is the macromolecule length. If one takes them into account, hydrodynamic interactions generically accelerate slithering. However if the plectoneme radius $R_0$ decreases, slithering is slower because the opposed strands move in opposite directions and $\tau_{\rm sl.,HI} \propto \eta L^3/\log(R_0/r_h)/k_{\rm B}T$, where $r_h$ denotes the polymer hydrodynamic radius~\cite{Marko1995,Toll2001}. In the DNA case, these relations lead to $\tau_{\rm sl.} \sim 100$~ms in water at room temperature when $L=3$~kbp. By contrast $\tau_{\rm sl.,HI} \sim 10$~ms for the same length and $R_0\simeq 5$~nm and $r_h=2$~nm. The latter time-scale is more consistent with experiments~\cite{Bussiek2002} and simulations, which indicates that hydrodynamic interactions indeed play a role, as expected. In the simulations of Ref.~\cite{Huang2001}, hydrodynamic interactions were taken into account through the Rotne-Prager tensor numerical scheme. Under physiological salt conditions, they found $\tau_{\rm sl.,HI} \simeq 3$~ms for $\sigma = -0.06$. 

In Ref.~\cite{Benedetti2014}, the authors also studied the complex formation by means of a mesoscopic numerical model. The protein-DNA complex capture time is given in Lennard-Jones units and coming back to real time units is uneasy because some computational tricks were used to accelerate the simulation. One can however estimate these times and they are in the 0.1 to 1~ms range, one order of magnitude faster than expected~\cite{Stasiak2016}. In the figure~6 of this reference, the capture times are plotted in function of the supercoiling density $\sigma$. A strong decrease is observed as $\sigma$ grows, in apparent contradiction with the logarithmic corrections discussed above. The explanation might come from the fact that the capture times were not measured at thermodynamical equilibrium, as indicated by the sentence ``although from time to time enhancer and promoter sites slither away resulting in very long off states (data not shown)''~\cite{Benedetti2014}. The observed short time-scales might also be related to this bias. They were measured in situations where the two sites do not wander too much away, whereas in the estimation of capture times discussed so far, the initial configuration was assumed to be random, both sites being separated by a distance on the order of the whole plectoneme length.

We have also seen above that the value of the torsion modulus $C$ is not entirely consensual. For this reason, we have made the same calculations with the alternative value $C=75$~nm. The numerical values of $R_0$, $P_0$, $k_p$ or $F(R_c)$ are only changed by few tens of percents at most, especially for the small values of $\sigma$, but the overall conclusions remain unchanged. 

In the future, we intend to go beyond the isotropic twistable WLC and to use the full Marko and Siggia elastic model~\cite{Marko1994,Carlon2017}. Even thought orders of magnitudes should be preserved, the twist-bend coupling ensuing from the difference between the minor and major grooves of DNA will likely lead to new interesting features of the system. The observed power laws for $k_p$ (figure~\ref{k}, left) also ought to be given an analytical explanation. 

\section*{Acknowledgements}

We thank Ivan Junier {and John H. Maddocks} for stimulating discussions and comments, as well as Catherine Tardin and Philippe Rousseau for their kind critical reading of the manuscript. 

\appendix

\section{{Frenet-Serret and relative torsions of a ribbon}}
\label{ribbon}

In differential geometry, a ribbon $(\mathbf{r},\mathbf{u})$ is defined by both a space curve $\mathbf{r}(s)$ and a unit vector $\mathbf{u}(s)$ perpendicular to the tangent vector $\mathbf{t}$ at each position $s$~\cite{Blaschke1950}. We have called $\mathbf{u}(s)$ the ribbon ``generatrix'' in the main text. The curve $\mathbf{r}(s)$ represents the DNA molecular axis and $\mathbf{u}(s)$ is in the same local plane as the base-pair ``rungs'' and thus locally defines the orientation of the double-helix material frame with respect to the Frenet-Serret frame (figure~\ref{ribbon:fig}). 

\subsection{Frenet-Serret or geometrical torsion of $\mathbf{r}(s)$}

\begin{figure}[ht]
\begin{center}
\includegraphics[height=6cm]{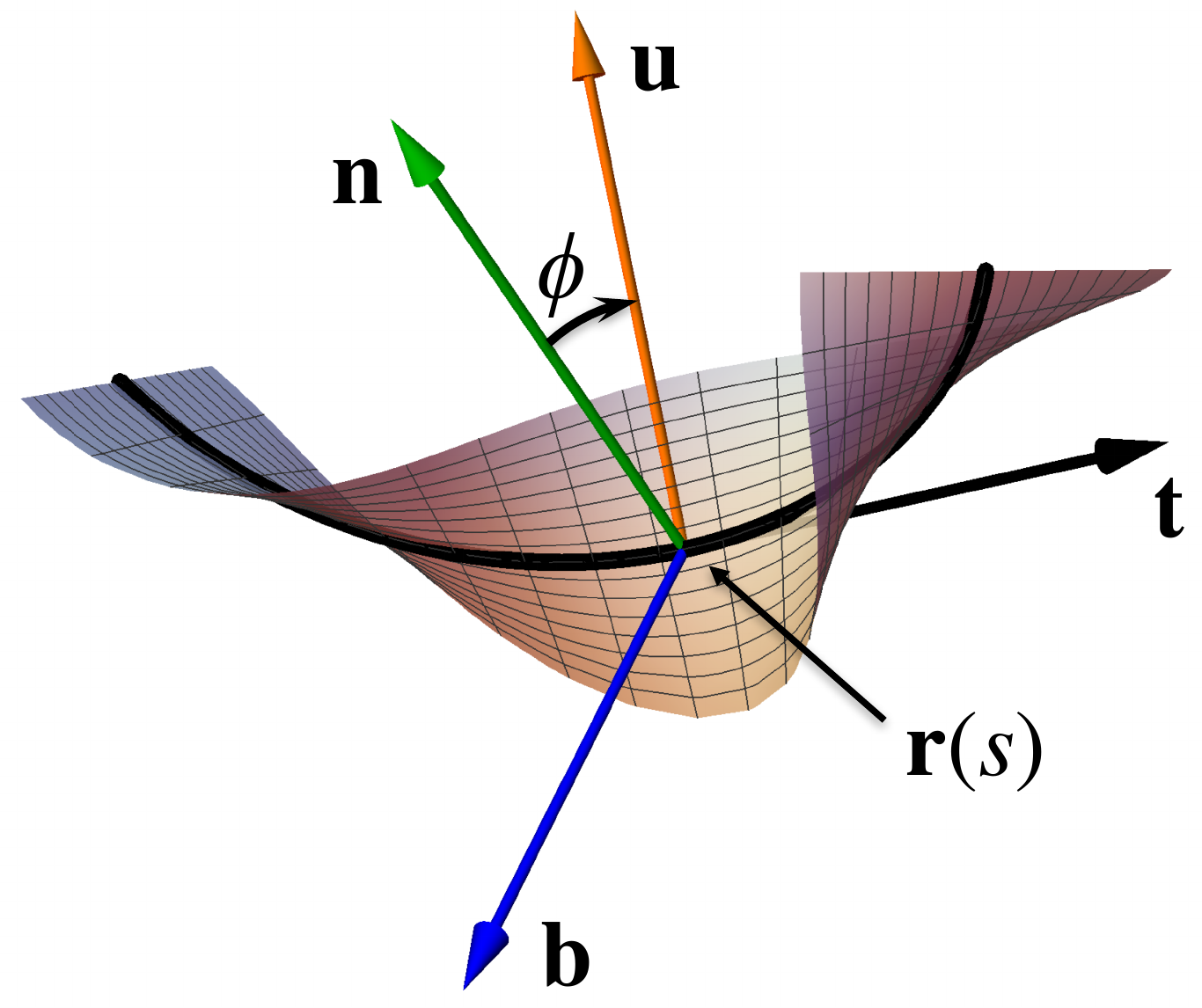} 
 \caption{A piece of ribbon defined by the curve $\mathbf{r}(s)$ (thick black line) representing the DNA molecular axis and the ``generatrix'' unit vector $\mathbf{u}(s)$ (in orange). We have also represented the local Frenet-Serret frame: the tangent vector $\mathbf{t}$ (black), the normal vector $\mathbf{n}$ (green) and the binormal vector $\mathbf{b}$ (blue). The vector $\mathbf{u}(s)$ belongs to the plane $(\mathbf{n},\mathbf{b})$ and makes an angle $\phi$ with $\mathbf{n}$. All these quantities depend on $s$.
 \label{ribbon:fig}}
 \end{center}
\end{figure}

To the curve $\mathbf{r}(s)$, we can indeed associate the local Frenet-Serret frame $(\mathbf{t}(s),\mathbf{n}(s),\mathbf{b}(s))$, as already discussed in the main text and as illustrated in figure~\ref{ribbon:fig}. The curve (or molecular axis) $\mathbf{r}(s)$ has a curvature $\gamma_{\rm axis}(s)$ and a torsion $\tau_{\rm axis}(s)$ that can be calculated through the Frenet-Serret formulae
\begin{eqnarray}
\frac{{\rm d} \mathbf{t}}{{\rm d}s} & = & \gamma_{\rm axis}(s) \mathbf{n}(s)\\
\frac{{\rm d} \mathbf{n}}{{\rm d}s} & = & - \gamma_{\rm axis}(s) \mathbf{t}(s) + \tau_{\rm axis}(s) \mathbf{b}(s)  \\
\frac{{\rm d} \mathbf{b}}{{\rm d}s} & = & -  \tau_{\rm axis}(s) \mathbf{n}(s)
\end{eqnarray}
This {\em Frenet-Serret} torsion $\tau_{\rm axis}(s)$  (or {\em Frenet-Serret} twist, up to a factor $2 \pi$) depends uniquely on the molecular axis $\mathbf{r}(s)$ and not on the generatrix $\mathbf{u}(s)$. 

\subsection{Relative torsion $a(s)$} 

By contrast, the torsion of the ribbon $(\mathbf{r},\mathbf{u})$ depends on both $\mathbf{r}(s)$ and $\mathbf{u}(s)$. It can be calculated from the formula~\cite{Klenin2000}
\begin{equation}
\tau_{\rm ribbon}(s) = \det\left[\mathbf{t}(s),\mathbf{u}(s),\frac{{\rm d}\mathbf{u}}{{\rm d}s}\right].
\end{equation}
If one introduces the angle $\phi(s)$ between $\mathbf{u}(s)$ and $\mathbf{n}(s)$ in the local plane $(\mathbf{n},\mathbf{b})$ (figure~\ref{ribbon:fig}), we can write $\mathbf{u}(s) = \cos (\phi) \mathbf{n} + \sin (\phi)  \mathbf{b}$. After a short calculation~\cite{Moffatt1992}, it follows that 
\begin{equation}
\tau_{\rm ribbon}(s) = \tau_{\rm axis}(s) + \frac{{\rm d} \phi}{{\rm d} s}.
\label{twists}
\end{equation}
The second term ${\rm d} \phi/{{\rm d} s}$, denoted by $a(s)$ in the main text, is what we call the {\em relative} torsion (or relative twist, up to a factor $2 \pi$). Contrary to $\tau_{\rm axis}(s)$, it depends on the choice of $\mathbf{u}(s)$, and comes in addition to $\tau_{\rm axis}(s)$. It measures how fast the DNA local frame winds around the Frenet-Serret frame along the curve. 

For a given curve $\mathbf{r}(s)$, its total writhe Wr is imposed because Wr only depends on $\mathbf{r}(s)$ and not on $\mathbf{u}(s)$. If in addition the ribbon is closed, its linking number Lk is fixed as a topological invariant. Thus owing to the Calugare\-anu-Fuller-White theorem~\cite{Marko2015,Vologodskii2015}, the total twist Tw is also fixed and $\int_{\bf r} \tau_{\rm ribbon}(s) {\rm d}s$ does not depend on the generatrix $\mathbf{u}(s)$.

In addition, $\int_{\bf r} \tau_{\rm axis}(s) {\rm d}s$ is constant for a given curve $\mathbf{r}(s)$. Owing to equation~\eqref{twists}, $\int_{\bf r} a(s) {\rm d}s$ is also constant. This could be anticipated since $\int_{\bf r} a(s) {\rm d}s = \int_{\bf r} ({\rm d} \phi/{{\rm d} s}) \, {\rm d}s$, which is a multiple of $2\pi$ for a closed ribbon. Consequently varying $\mathbf{u}(s)$ at fixed molecular axis shape $\mathbf{r}(s)$ and fixed ribbon topology keeps $\int_{\bf r} a(s) {\rm d}s$ constant.

\section{{Quadratic forms $Q_b(s)$, $Q_t(s)$ and $Q_c(s)$}}
\label{quads}

From equation~\eqref{C2}, we get
\begin{eqnarray}
Q_b(s) & = & \frac{l_p}2 \left\{\left[ r_1''(s) - \Omega_0^2 r_1(s) -2 R_0 \Omega_0 \theta_1'(s) \right]^2 
	+ \phantom{\frac12}\right. \\
 & + & \left[2 \Omega_0 r_1'(s) + R_0 \theta_1''(s) \right]^2 + 
 	\frac{\Omega_0^2 R_0^2}{1-\Omega_0^2 R_0^2} [\Omega_0 r_1'(s)+R_0 \theta_1''(s)]^2 \\
 & + &  \left. 2 R_0^2 \Omega_0^2 \theta_1'(s)^2 + 4 R_0 \Omega_0^3 r_1(s) \theta_1'(s)\phantom{\frac12} \right\}
 \label{Qb}
 \end{eqnarray}
 for bending. As for confinement,
\begin{eqnarray}
Q_c(s) & = & \frac{5}9 \frac{K}{R_0^{8/3}} r_1(s)^2 
\label{Qc}
\end{eqnarray}
by developing the confinement energy density $K/R(s)^{2/3}$ at order 2 in $\varepsilon$.

The torsional energy density is
\begin{equation}
\frac{C}2 \tau(s)^2  = \frac{C}2 \left[ \sigma \omega_0 + \Delta \tau_0 + \varepsilon \tau_1(s) + \varepsilon^2 \mathcal{T}_1(s)+ \varepsilon^2 \tau_2(s) + \mathcal{O}(\varepsilon^3)  \right]^2.
\end{equation}
Here we have come back to the original ribbon \emph{with} intrinsic supercoiling $\sigma \omega_0$ and we have explicitly made the distinction between the order-2 terms of $\tau(s)$ {\em linear} in $r_2(s)$, $\theta_2(s)$, $a_2(s)$, and their derivatives, that we have grouped in $\tau_2(s)$; and the order-2 terms {\em quadratic} in $ r_1(s)$, $ \theta_1(s)$ and their derivatives, coming from the Frenet-Serret torsion, and grouped in $\mathcal{T}_1(s)$.
The total order-2 {\em quadratic} contribution to the twisting energy density is thus 
\begin{eqnarray}
Q_t(s) & = & \frac{C}2 \left[  \tau_1(s)^2 + 2 (\sigma \omega_0 + \Delta \tau_0) \mathcal{T}_1(s)  \right]
\label{Qt}
\end{eqnarray}
Using again the Frenet-Serret relation $\tau = - {\rm d} \mathbf{b} / {\rm d}s \cdot \mathbf{n}$ and developing it at order $\varepsilon^2$, one gets
\begin{equation}
\mathcal{T}_1(s) = \frac{\mathcal{N}(s)}{2 R_0^2 \Omega_0^4 \left(1-R_0^2 \Omega_0^2\right)^{3/2}}
\end{equation}
with 
\begin{eqnarray}
\mathcal{N}(s) & = & -2 R_0^6 \Omega_0^7 \theta_1'(s)^2+6 R_0^4 \Omega_0^3 \theta_1''(s)^2+3 R_0^4 \Omega_0^5\theta_1'(s)^2-2 R_0^3 \Omega_0^2 r_1^{(3)}(s) \theta_1''(s)+6 R_0^3 \Omega_0^4 r_1''(s) \theta_1'(s) \nonumber \\
  & + & 2  R_0 \theta_1^{(3)}(s) \left\{R_0 \Omega_0 \left[\left(3
   R_0^2 \Omega_0^2-2\right) \theta_1'(s)-R_0 \Omega_0 r_1''(s)\right]+r_1''(s)\right\} \nonumber \\
   & + & 2 \Omega_0^3 r_1(s)
   \left\{R_0^2 \Omega_0 \left[R_0 \Omega_0^2 \left(3-2
   R_0^2 \Omega_0^2\right) \theta_1'(s)+R_0\theta_1^{(3)}(s)+3 \Omega_0 r_1''(s)\right]-2 r_1''(s)\right\} \nonumber \\
   & - & 4 R_0^2 \Omega_0^3 r_1''(s)^2+R_0^2 \Omega_0^7
   r_1(s)^2-4 R_0^2 \Omega_0 \theta_1''(s)^2 + \Omega_0^3 \left(-3 R_0^4 \Omega_0^4+9 R_0^2 \Omega_0^2-4\right) r_1'(s)^2 \nonumber \\
   & + & 2 \Omega_0 r_1'(s) \left[-2 \left(R_0^2 \Omega_0^2-1\right) r_1^{(3)}(s)-R_0
   \Omega_0 \left(R_0^4 \Omega_0^4-5 R_0^2 \Omega_0^2+2\right) \theta_1''(s)\right] \nonumber \\
   & + & 2 R_0 r_1^{(3)}(s) \theta_1''(s)-4 R_0 \Omega_0^2 r_1''(s)\theta_1'(s)+4 \Omega_0 r_1''(s)^2.
\label{N:s}
\end{eqnarray}

\section{Full expression of the Hermitian matrix $M(q)$}
\label{full:M}

By using Parseval's theorem, we can write $M(q)=M_b(q)+M_t(q)+M_c(q)$, where the three terms are:
\begin{enumerate}
\item The contribution of bending coming from equation~\eqref{Qb}:
\begin{equation}
M_b(q) = \frac{\ell_p}{a-1} \left(
\begin{array}{ccc}
\Omega_0^4(a-1) - q^2 \Omega_0^2(2-a)+q^4(a-1) & R_0\Omega_0 \left[4 \Omega_0^2(a-1) - a q^2 \right] & 0 \\
R_0\Omega_0 \left[4 \Omega_0^2(a-1) - a q^2 \right] & R_0^2\left[ 6 \Omega_0^2(a-1) - q^2\right]  & 0 \\
0 & 0 & 0
\end{array}
\right)
\end{equation}
where $a=R_0^2\Omega_0^2$;
\item The contribution of torsion deriving from equations~(\ref{Qt}--\ref{N:s}):
\begin{equation}
M_t(q) = C \, Y(q) Y(q)^* + 2 C (\sigma \omega_0 + \Delta \tau_0) N(q)
\end{equation}
where
\begin{equation}
Y(q) = 
\left(
\begin{array}{c}
\frac1{\sqrt{1-a}} \frac{\Omega_0}{R_0} \left[-a + (2-a)\frac{q^2}{\Omega_0^2}   \right]\\
\frac1{\sqrt{1-a}} \left[1-2a + \frac{q^2}{\Omega_0^2}\right] \\
1 
\end{array}
\right)
\end{equation}
and
\begin{equation}
N(q) = \frac{1}{2\Omega_0^2 a (1-a)^{3/2}} 
\left(
\begin{array}{ccc}
\Omega_0^3 \left[ \Omega_0^2 a + 3a (1-a)q^2 \right] & R_0 \Omega_0^2 a \left[ \Omega_0^2(3-2a) + (1-a) q^2 \right] & 0 \\
R_0 \Omega_0^2 a \left[ \Omega_0^2(3-2a) + (1-a) q^2 \right] & \Omega_0 a^2 (3-2a) & 0 \\
0 & 0 & 0 
\end{array}
\right)
\end{equation}
represents $\mathcal{N}(s)$ in the Fourier space;
\item And the contribution of confinement ensuing from equation~\eqref{Qc}:
\begin{equation}
M_c(q) = \frac{10}9 \, \frac{c}{\ell_p^{1/3}R_0^{8/3}}
\left(
\begin{array}{ccc}
1 & 0 & 0 \\
0 & 0 & 0 \\
0 & 0 & 0
\end{array}
\right).
\end{equation}
\end{enumerate}

\end{document}